\documentclass[twocolumn,showpacs,preprintnumbers,amsmath,amssymb]{revtex4}
\usepackage{graphicx}% Include figure files
\usepackage{dcolumn}% Align table columns on decimal point
\usepackage{bm}% bold math
\usepackage{upgreek}

\begin{document}

%\preprint{APS/123-QED}

\title{A metastable superconducting qubit}

\author{Andrew J. Kerman}
\affiliation{Lincoln Laboratory, Massachusetts Institute of
Technology, Lexington, MA, 02420}

\date{\today}

\begin{abstract}
We propose a superconducting qubit design, based on a tunable
RF-SQUID and nanowire kinetic inductors, which has a dramatically
reduced transverse electromagnetic coupling to its environment, so
that its excited state should be metastable. If electromagnetic
interactions are in fact responsible for the current excited-state
decay rates of superconducting qubits, this design should result in
a qubit lifetime orders of magnitude longer than currently possible.
Furthermore, since accurate manipulation and readout of
superconducting qubits is currently limited by spontaneous decay,
much higher fidelities may be realizable with this design.
\end{abstract}

\maketitle

One of the distinguishing features of Josephson-junction (JJ)-based
qubits is their strong coupling to electromagnetic (EM) fields,
which permits fast gate operations ($\sim$ 10-100ns). However, it
may also be responsible for their short excited-state lifetimes
($\lesssim 4\mu$s \cite{onef,flux photon, transmon expt,good
phase,quantronium}); that is, assuming the decay process is
electromagnetic, its rate depends on the same matrix element which
governs intentional qubit manipulations by external fields.
Unfortunately, understanding and controlling spontaneous decay of
these circuits has so far proved difficult, because it also depends
on their EM environment at GHz frequencies. This environment is
affected not only by packaging and control lines, but also by
microscopic degrees of freedom in the substrate, surface oxides, and
JJ barrier dielectrics. In fact, low-frequency noise due to
microscopic fluctuators is already known to produce ``dephasing" of
qubits \cite{onef,flux photon,transmon expt,good phase,quantronium}.
Although little is yet certain about the properties of these degrees
of freedom, work is ongoing to study them \cite{TLSs}, and to reduce
their number through improved materials and fabrication
\cite{materials}.

In this Letter we discuss a different approach, seeking a qubit
which is insensitive to high-frequency EM fluctuations \textit{by
design}. This is a departure from the research area known as circuit
QED \cite{circuit QED}, in which strong transverse coupling to EM
fields is both a prerequisite and a figure of merit. We will show
that a qubit design based on weak transverse EM coupling could yield
significantly longer excited-state lifetimes, while still allowing
practical, scalable computation.

\begin{figure*}
\includegraphics[width=7in]{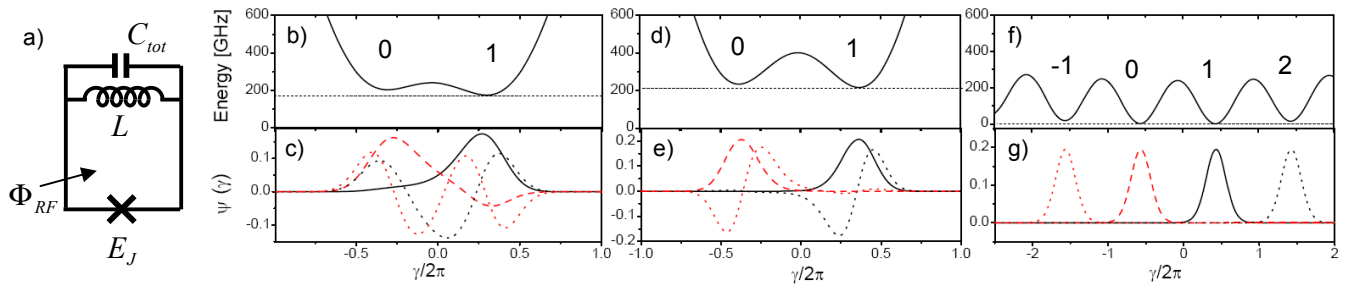}
\caption{\label{fig:1} (color online) Fluxon tunneling for the RF
SQUID flux qubit. (a) schematic. (b) potential, and (c) qubit level
wavefunctions $|g\rangle$ (solid) and $|e\rangle$ (dashed) at
$f=0.515$ for $E_J,E_C,E_L=h\times 120,6,60$ GHz. (d) potential, and
(e) wavefunctions, for $E_J,E_C,E_L=h\times 180,4,60$; the potential
barrier between wells is higher so the tunneling is weak. (f)
potential, and (g) wavefunctions for $E_J,E_C,E_L=h\times
120,6,0.375$. Dotted lines in (c), (e) and (g) show the next excited
states; In (c) and (e) these are ``vibrational" excitations, while
in (g) they are the ground states of adjacent wells (corresponding
to -1 or 2 fluxons in the SQUID loop; fluxon numbers for each well
are indicated in (f)).}
\end{figure*}

The decay rate of an excited state $|e\rangle$ to lower-lying state
$|g\rangle$ is typically given by Fermi's golden rule: $\Gamma\equiv
1/T_1=(2\pi/\hbar)|m_i|^2\rho(\hbar\omega_{eg})$, where
$m_i\equiv\langle e|\hat{H}_i|g\rangle$, $\hat{H}_i$ is a
Hamiltonian describing the coupling between the qubit and a
continuum (e.g., the excited states of an ensemble of two-level
systems - TLSs), and $\rho(\hbar\omega_{eg})$ is the density of
states in that continuum at the energy $\hbar\omega_{eg}\equiv
E_e-E_g$. The $m_i$ can be nonzero for a JJ-based qubit when the
flux through a loop, the induced charge across a JJ, or a JJ
critical current depends on the state of one or more TLSs. To
minimize the resulting decay rate, we must reduce $\rho$ or $m_i$.
Our focus here will be on the latter.

A good choice for qubit energy levels which are weakly coupled to
each other by EM fields are the flux states of an RF SQUID [Fig.
\ref{fig:1}(a)] \cite{lukens} at large $E_J/E_C$, where
$E_J=\Phi_0I_C/2\pi$ and $E_C\equiv e^2/2C_{tot}$ are the Josephson
and charging energies ($I_C$ is the JJ critical current,
$\Phi_0\equiv h/2e$, and $C_{tot}$ is the total capacitance across
the JJ). When $\Phi_{RF}\sim\Phi_0/2$, two quantum states, in which
either zero or one fluxon is contained in the loop, become nearly
degenerate, and separated by a potential barrier [Fig.
\ref{fig:1}(b)]. The Hamiltonian for the RF SQUID is \cite{inductive
shunt}:

\begin{equation}
\hat{H}=4E_C(\hat{n}-n_e)^2-E_J\cos\hat\phi+E_L\hat\gamma^2/2\label{eq:hamiltonian}
\end{equation}

\noindent where $\hat\phi$ is the phase across the JJ,
$\hat{n}\equiv -id/d\phi$ is operator corresponding to the number of
Cooper pairs that have tunneled through the JJ,
$\hat\gamma\equiv\hat\phi+2\pi f$ is the phase across the inductor,
$f=\Phi_{RF}/\Phi_0$, and $E_L\equiv(\Phi_0/2\pi)^2/L$. The quantity
$n_e$ is a fluctuating offset charge across $C_{tot}$ induced by
capacitances to the environment or by tunneling of quasiparticles
through the junction (at DC $n_e$=0 due to the inductive shunt).

We diagonalize $\hat{H}$ on a lattice of $\gamma$ points to obtain
wavefunctions $\psi_k(\gamma)\equiv\langle\gamma|k\rangle$ [Fig.
\ref{fig:1}(c), (e), and (g)], which are then used to evaulate
transition matrix elements $\langle k|\hat{H}_i|k^\prime\rangle$
\cite{nori} for flux, charge, and $I_C$-coupled TLSs, with:
$\hat{H}_f\equiv 2\pi\delta fE_J\sin(\hat\gamma+2\pi f)$,
$\hat{H}_n\equiv 8\delta nE_C\hat{n}$, and $\hat{H}_I\equiv\delta
I_CE_J\cos(\hat\gamma+2\pi f)$, respectively, and $\delta f$,
$\delta n$, and $\delta I_C$ are the (small) amplitudes of
TLS-state-dependent changes in $f$, $n_e$, and $I_C$. These
amplitudes will be different for each TLS, so it is conceptually
useful to recast the golden rule in terms of an average noise power
spectral density $S_i$ \cite{nori} thus: $\Gamma_i\equiv
|d_i|^2S_i(\omega_{eg})/\hbar^2$, where $d_i\equiv\langle
e|\hat{X}_i|g\rangle$ are analogous to a transition dipole for each
fluctuation, and: $\hat{X}_f\equiv 2\pi E_J\sin(\hat\gamma+2\pi f)$,
$\hat{X}_n\equiv 8E_C\hat{n}$, $\hat{X}_{I_C}\equiv
E_J\cos(\hat\gamma+2\pi f)$ with units of energy per $\Phi_0$,
electron pair, and current.

Since the operators $\hat{X}_i$ are local in $\hat\gamma$, a simple
way to reduce all of the $d_i$ at once is to reduce the overlap of
the probability distributions $|\psi_g(\gamma)|^2$ and
$|\psi_e(\gamma)|^2$. This overlap results from tunneling through
the barrier [Fig. \ref{fig:1}(b),(c)], so to minimize it we detune
the left and right wells from each other ($f\neq 0.5$) to avoid
resonant tunneling, and increase the barrier height by increasing
$E_J/E_C$ and/or $E_J/E_L$ [Fig. \ref{fig:1}(d),(e)].

Unfortunately, when $f\neq 0.5$, $d\omega_{eg}/df\neq 0$, and
nonzero, low-frequency $\delta f$ produce dephasing \cite{onef}.
This sensitivity can be reduced by increasing $L$, since
$\hbar\omega_{eg}\approx \frac{\Phi_0^2}{L}(f-0.5)$, for $E_L\ll
E_J$ \cite{lukensEJ}. To realize large $L$, increasing the loop size
is not attractive, both because it would need to be of millimeter
scale, and because its large capacitance would limit $E_C$. Instead,
we propose using the kinetic inductance of a long, meandered
nanowire patterned from thin ($\sim$5 nm thick) NbN, which can have
sheet inductance as large as $\sim$100 pH and $I_C\sim 20\mu$A
\cite{kinetic inductance,JJ array}. A 10 $\mu$m-square meander of
100 nm-wide wire gives $L\sim$500 nH \cite{kinetic inductance}, and
EM simulation shows a shunt capacitance of only $\sim$0.4 fF,
significantly smaller than that of the JJs we consider below
($\sim$3.2 fF).

\begin{figure}
\includegraphics[width=3.25in]{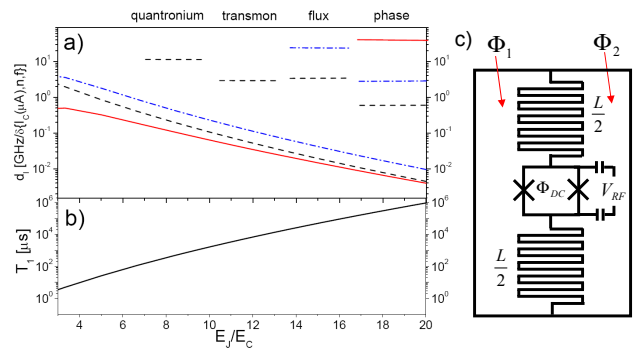}
\caption{\label{fig:transverse}(color online) Transverse coupling of
the metastable RF SQUID qubit vs. $E_J/E_C$, for $E_C,E_L=h\times
6,0.375$ GHz, and $f=0.57$. Panel (a) shows $|\langle
e|X_n|g\rangle|$ (dashed), $|\langle e|X_f|g\rangle|$ (dash-dot),
and $|\langle e|X_I|g\rangle|$ (solid), respectively. Horizontal
lines show equivalent $|\langle e|X_i|g\rangle|$ for the transmon
\cite{transmon expt}, quantronium \cite{quantronium}, flux
\cite{onef,flux photon}, and phase \cite{good phase} qubits. (b)
shows the predicted excited state lifetime for the metastable RF
SQUID qubit. (c) schematic for the proposed qubit, with $E_J/E_C$
tunable through the flux $\Phi_{DC}$. The gradiometric design
decouples this flux from $\epsilon=I_p(\Phi_1-\Phi_2)$
\cite{twoloop}.}
\end{figure}

Figure \ref{fig:transverse}(a) shows the resulting $|d_i|$ for our
proposed qubit, as a function of $E_J/E_C$. Also shown, by
horizontal dashed lines, are the $|d_i|$ for transmon, quantronium,
flux, and phase qubits \cite{charge basis}. Based on these results,
and by extracting bounds on the $S_i$ from $T_1$ values observed in
Refs. \cite{transmon expt,good phase,onef,flux photon,quantronium},
we can estimate $T_1$ for our qubit. Not surprisingly, no single set
of $S_i$, in conjunction with the calculated $d_i$, can accurately
explain all of the observations, since the noise levels are likely
somewhat different in each experiment; however, for the present purpose, we take: $S_I(5
\textrm{GHz})\lesssim 1.4\times 10^{-17}\mu$A$^2$Hz$^{-1}$, from
$T_1=650$ns for the phase qubit of Ref. \cite{good phase}; $S_n(5.7
\textrm{GHz})\lesssim 1.6\times 10^{-15}$ Hz$^{-1}$ from $T_1=1.7
\mu$s for the transmon of Ref. \cite{transmon expt}; and $S_f(5.5
\textrm{GHz})\lesssim 1.3\times 10^{-20}$ Hz$^{-1}$ from $T_1=2\mu$s
from the flux qubit of Ref. \cite{onef}. Panel (b) shows the
resulting estimate of $T_1$ for our qubit (dominated by charge
noise). For $E_J/E_C\sim 3$ (as in Ref. \cite{fluxonium}), $T_1\sim
3\mu$s, similar to what was observed; however, at $E_J/E_C=20$, we
find $T_1\sim$ 950 milliseconds \cite{phonons}.

The reduced transverse coupling that we achieve through increasing
$E_J/E_C$ also means we must drive the qubit with larger fields to
manipulate it \cite{Raman}, or measure it dispersively \cite{JJ
array,weak coupling}. At some point the required driving becomes
strong enough that spurious effects occur, such as off-resonant
excitation of strong transitions to short-lived excited states
(followed by decay), or large nonlinearities in the qubit response.
Furthermore, initializing the qubit state will take longer as the
lifetime is increased \cite{cooling}. It will therefore be useful to
be able to adjust $E_J/E_C$ in real time using a tunable RF SQUID
[Fig. \ref{fig:transverse}(c)] (analogous to the tunable flux qubit
\cite{orlando,twoloop}). The single JJ is replaced by a DC SQUID,
and the RF SQUID loop is replaced with a gradiometric design where
$f_{RF}\equiv (\Phi_1-\Phi_2)/\Phi_0$. In this configuration, $E_J$
in eq. \ref{eq:hamiltonian} is replaced with:

\begin{eqnarray}
E_J(f_{DC})&=&2E_{J0}\cos[\pi f_{DC}-\hat\gamma_{DC}/2]\label{eq:EJ}\\
& \approx & 2E_{J0}^\prime\cos[\pi f_{DC}]\label{eq:approxEJ}
\end{eqnarray}

\noindent where $E_{J0}$ is the Josephson energy of each JJ,
$f_{DC}=\Phi_{DC}/\Phi_0$, and $\hat\gamma_{DC}$ is the phase across
$L_{DC}$, the self-inductance of the DC SQUID loop. To obtain eq.
\ref{eq:approxEJ}, we note that for $L_{DC}\ll
L,L_J\equiv\Phi_0/2\pi I_C$, the zero-point fluctuations of
$\hat\gamma_{DC}$ can be adiabatically eliminated, yielding to
leading order only a small renormalization of $E_{J0}$ \cite{self
ind,self ind 2} (for $L_{DC}<50$ pH, and the parameters under
consideration here, a fraction of a percent).

%the zero-point motion of
%$\hat\gamma_{DC}$ is governed by the Hamiltonian:
%$\hat{H}_{DC}=4E_C^{DC}(\hat{n}_1-\hat{n}_2)^2+E_L^{DC}\hat\gamma_{DC}^2/2$,
%where $E_C^{DC}$ is the charging energy of each JJ, $i=1,2$ denote
%the two JJs, and $E_L^{DC}=(\Phi_0/2\pi)^2/L_{DC}$. 

The qubit can be manipulated (or measured dispersively) with
$V_{RF}$, $\Phi_{DC}$, or $\Phi_{RF}\equiv\Phi_1-\Phi_2$  [Fig.
\ref{fig:transverse}(c)]. We consider the first two here. In order
to describe large-amplitude driving, and to incorporate spontaneous
decay between instantaneous energy eigenstates
$|m(t)\rangle^\prime$, we use a time-dependent transformation to the
instantaneous energy eigenbasis, yielding the Hamiltonian:
$\hat{H}_{ad}=\hat{R}\hat{H}\hat{R}^\dagger-i\hbar\hat{R}\frac{d}{dt}\hat{R}^\dagger$
where $\hat{H}$ is given by eq. \ref{eq:hamiltonian} and $\hat{R}$
is defined by: $\hat{R}|\psi\rangle\equiv|\psi^\prime\rangle$ (the
prime indicates the basis $|m(t)\rangle^\prime$). The first term is
diagonal, containing the time-dependent eigenenergies, and the
second term yields transitions between levels. We integrate a master
equation based on $\hat{H}_{ad}$, truncated to the 10 lowest-lying
instantaneous eigenstates (up to $\gtrsim$100 GHz above
$|g\rangle$). To this we add a spontaneous decay rate
$\gamma_{mn}(t)$ from each level $|m^\prime\rangle$ to each other
level $|n^\prime\rangle$. To generate the $\gamma_{mn}(t)$, we use
Fermi's golden rule, and assume an Ohmic noise spectrum
$S_i(\omega)\propto\hbar\omega/(1-e^{-\hbar\omega/k_BT})$
($\omega>0$ denotes downward transitions, $\omega<0$ upward), with
the overall amplitude for each type of noise discussed above. The
time dependence of the $\gamma_{mn}(t)$ comes from the
$|X_i(t)|_{mn}^2$.

\begin{figure}
\includegraphics[width=3.25in]{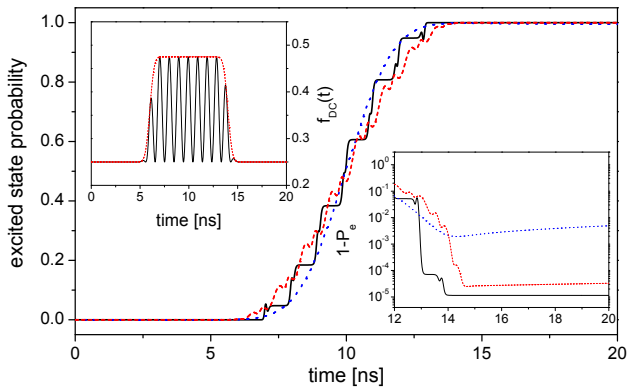}
\caption{\label{fig:master} (color online) Transverse manipulation
of the metastable RF SQUID qubit. Integration of the Master equation
(including decay) for the first 10 levels of the RF SQUID,
undergoing a $\pi$-pulse starting from $|g\rangle$. The solid line
is for the modulation of $\Phi_{DC}$ in the left inset; the dashed
line is for a sinusoidal modulation of $V_{RF}$ with amplitude
$n_e=0.24$, at $E_J=h\times$42 GHz. The dotted line is the
equivalent result for a flux qubit \cite{onef,flux photon}.}
\end{figure}

As a test case, we consider a $\pi$-pulse, where the qubit starts in
$|g\rangle$, for which an indication of gate fidelity is how much
population we can put in $|e\rangle$, as shown in Fig.
\ref{fig:master}. We take $E_C,E_L=h\times 6,0.375$ GHz ($L=430$
nH), and $f_{RF}=0.57$ ($\omega_{eg}=2\pi\times$1.034 GHz
\cite{detuning}). For modulation of $\Phi_{DC}$ (solid line), we use
the pulse shown in the left inset to Fig. \ref{fig:master}, which
starts and ends at $E_J=h\times 200$GHz (with
$E_{J0}^\prime=h\times$280 GHz). For modulation of $V_{RF}$, we take
a fixed $E_J=h\times$42 GHz. The simulation yields $1-P_e=1.1\times
10^{-5}$ and $1-P_e=2.5\times 10^{-5}$ for $\Phi_{RF}$ and $V_{RF}$
modulation, respectively. The former is limited almost completely by
decay of $|e\rangle$ during the brief excursions to smaller
$E_J/E_C$ where $\gamma_{10}(t)$ is larger. This also explains the
shape of the time evolution: the drive becomes effectively faster
when $E_J/E_C$ is smaller, producing the upward ``steps". Spurious
excitation to adjacent fluxon states (-1,2 in Fig. \ref{fig:1}(f))
and higher vibrational states is at the $\mathcal{O}(10^{-6})$
level. Driving with $V_{RF}$ is limited by off-resonant excitation
of the first vibrational levels (at $\sim$40 GHz) followed by decay.
This process is suppressed for $\Phi_{DC}$ modulation since the
perturbation is nearly even about the potential well center. For
comparison is shown the same simulation for a flux qubit
\cite{onef,flux photon}, which has $1-P_e=2\times 10^{-3}$, due to
decay from $|e\rangle$.

This simulation does not include 1/f flux noise \cite{onef}. To
estimate its effect, we use the results of Ref. \cite{ithier}, and
the fact that for $L=430$ nH, $d\omega_{eg}/df=14.3$ MHz/m$\Phi_0$
($\sim$100 times smaller than a typical flux qubit far from
$f=0.5$). For the noise amplitude measured in Ref. \cite{onef}, we
calculate the average error in the qubit relative phase over the 8
ns $\pi$-pulse to be $\sim 4.5$ mrad, which for the maximally
sensitive $(|g\rangle+|e\rangle)/\sqrt{2}$ state gives an error
probability of only $\sim 2.0\times 10^{-5}$ \cite{long term}.

\begin{figure}
\includegraphics[width=3.25in]{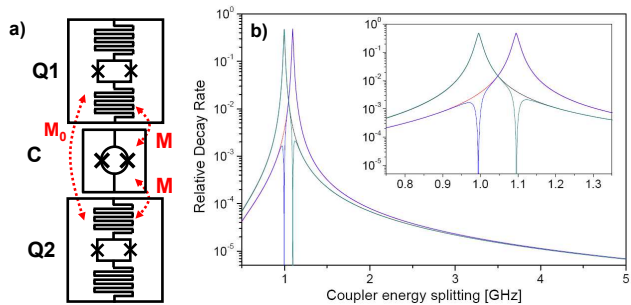}
\caption{\label{fig:coupled} (color online) Switchable coupling
between metastable RF SQUID qubits. (a) schematic; two data qubits
(RF SQUIDs with meandered kinetic inductors) are coupled through a
mutual inductance $M$ to a coupler qubit (RF SQUID with large $I_p$,
and $\epsilon_C=0$) and parasitically to each other ($M_0$). (b)
shows the calculated decay rates for the four computational levels,
relative to the decay rate of the coupler in isolation, for the
parameters in the text, and with $\epsilon_1,\epsilon_2=h\times
1.0,1.1$GHz (chosen to be different only for clarity). The inset shows the resonances that occur due to
nonzero $\delta_i$.}
\end{figure}

By eliminating the transverse coupling between qubit levels induced
by external fields, we have also eliminated the usual mechanism for
coupling qubits to each other \cite{transverse}. Instead, we must
use a longitudinal coupling, similar to Ref. \cite{simulation},
which makes use of the nonzero flux tunability \cite{semba}. A
schematic of our proposed circuit is shown in Fig.
\ref{fig:coupled}(a). Two RF SQUID qubits are coupled by mutual
inductances $M$ to a third coupler qubit with large persistent
current $I_p^C$, biased at its degeneracy point ($f_{RF}^C=0.5$).
The approximate Hamiltonian is:

\begin{equation}
\hat{H}\approx\sum_i^{1,2,C}
[\epsilon_i\hat\sigma_i^z+t_i\hat\sigma_i^x]+
J_C\hat\sigma^z_C\sum_j^{1,2}[\hat\sigma^z_j+\delta_j\hat\sigma^x_j]+J_0\hat\sigma^z_1\sigma^z_2\label{eq:coupled}
\end{equation}

\noindent Here, eigenstates of $\hat\sigma^z$ are well-defined flux
(persistent current) states, and $J_C=MI^C_pd\epsilon_{1,2}/d\Phi$
with $d\epsilon_{1,2}/d\Phi\approx 4\pi^2 E_L/\Phi_0$
\cite{simulation}, $J_0=M_0I_p^1I_p^2$. The $\delta_j$ are residual
transverse flux coupling of the data qubits due to nonzero $d_f$. We
take: $I^C_p=5.2\mu$A and $t_C=h\times 5$GHz \cite{coupler},
$M=5$pH, $M_0=0.1$pH, $L=430$nH, and $I_p^{1,2}\approx
\Phi_0/2L=2.4$ nA, to obtain $J_C=h\times$188 MHz, $J_0=h\times
0.87$ kHz; this gives a conditional frequency shift
$h\delta_\nu\approx 2J_C^2/t_C-J_0=14.1$ MHz \cite{simulation} and a
conditional-$\pi/2$ gate in $\approx$ 18 ns. If we use spin-echoes
\cite{ithier,simulation} during the gate, the residual phase drift
due to 1/f flux noise (not canceled by the echo) during this time is
3.6 mrad, producing a maximal error (in addition to that from the
$\pi$ pulse) of $3.0\times 10^{-6}$ \cite{long term}.

A very small transverse coupling to the data qubits ($\delta_j\ll
1$) also means that their excited states will undergo negligible
mixing with the excited state of the coupler (which will likely be
short-lived). Figure \ref{fig:coupled}(b) shows the decay rates that
result. These are proportional to: $|\langle
ij|\hat\sigma^z_C|kl\rangle|^2$ $(i,j,k,l\in \{g,e\})$, where
$|ij\rangle$  are the computational states (the lowest four
eigenstates of eq. \ref{eq:coupled}, which in the $J_C\rightarrow 0$
limit correspond to the coupler in its ground state
\cite{simulation}). The pronounced peaks (and dips) occur when the
coupler is nearly resonant with one of the data qubits; in these
regions, the nonzero $\delta_j$ produce two entangled states of a
data qubit and the coupler, with one state coupling maximally to
fluctuations and the other minimally. When both qubits are detuned
far from the coupler, the data qubit decay rate is sufficiently
suppressed that even coupler qubit lifetimes at the ns scale would
have little effect.

In summary, we have described a qubit design with weak transverse
coupling to EM fields. This qubit should be significantly less
sensitive to microscopic EM degrees of freedom arising from
fabrication imperfections, and may permit very long $T_1$ times with
good fabrication yield. Although these qubits are still weakly
sensitive to low-frequency flux drifts, this is in principle a
problem with any manufactured qubit; e.g., even a flux qubit at its
flux-insensitive point is sensitive to drifts of $I_C$ at a level
that may soon become a coherence limit \cite{ICvar}, and even
lithographically defined superconducting resonators have
resonance-frequency noise \cite{resonators}. Thus, man-made qubits
may inevitably require modified encoding/computation schemes which
are resistant to the inevitable drift between each qubit's relative
phase and the absolute phase reference required for sustained
computation.

We acknowledge helpful discussions with William Oliver, Jeremy Sage,
and Jens Koch.

This work is sponsored by the United States Air Force under Contract
\#FA8721-05-C-0002. Opinions, interpretations, recommendations and
conclusions are those of the authors and are not necessarily
endorsed by the United States Government.


\begin{thebibliography}{}
\bibitem{JJ array} An alternative is a series array of
JJs, as demonstrated in Ref. \cite{fluxonium}. However, since each
of these JJs is a tunneling path for the fluxon, this approach
cannot be used to strongly suppress $d_i$ in the manner discussed here. Although fluxons can in
principle tunnel through NbN nanowires, due to their granular
microstructure or quantum phase slips, we estimate these processes
occur at rates $\lesssim\mathcal{O}$(10 s$^{-1}$).
\bibitem{lukensEJ} Note that Ref. \cite{lukens} was in the opposite limit $E_J\ll
E_L$.
\bibitem{charge basis} For the transmon and quantronium, the $T_1$ is only sensitive to
charge noise, and for the flux qubit at $f=0.5$, charge and flux
noise. For all three, we diagonalize the Hamiltonian in the charge
basis.
\bibitem{weak coupling} Weak coupling (large $E_J/E_C$) may also be
interesting for dispersive readout. In contrast to the
strong-coupling limit, in which dephasing due to photon number
splitting is important \cite{flux photon,photon
dephasing,quantronium}, for weak coupling large fields could be
used, with a corresponding reduction in (fractional) photon-number
uncertainty.
\bibitem{detuning} A relatively small $\omega_{eg}$ is used to
avoid spurious excitation of the adjacent fluxon states by a short
pulse (at $L=430$nH these states are only $\sim 15$ GHz away).
\bibitem{long term} These phase drifts continue to accrue with time,
even when spin-echoes are used (e.g., eq. 22 of Ref. \cite{ithier}).
For the present qubit, this is $(13 \mu\textrm{s})^{-1}$ for the
flux noise measured in Ref \cite{onef}.
\bibitem{semba} A related proposal was detailed in [Y.-D. Wang, A. Kemp, and K. Semba, Phys. Rev. B
{\bf 79} 024502 (2009)]. In this scheme the coupling is switched off
by tuning the qubits to denegeracy, so it would not be applicable
here.
\end{thebibliography}
\end{document}